\documentclass{article}

\usepackage{arxiv}

\usepackage[utf8]{inputenc} 
\usepackage[T1]{fontenc}    

\usepackage[hidelinks]{hyperref}

\usepackage{url}            
\usepackage{booktabs}       
\usepackage{nicefrac}       
\usepackage{microtype}      
\usepackage{lipsum}		
\usepackage{subcaption}
\usepackage{graphicx}
\usepackage[numbers]{natbib}
\usepackage{doi}

\usepackage{booktabs} 
\usepackage{floatrow}

\usepackage{amsmath,amssymb,amsfonts}

\usepackage{authblk}

\title{Near out-of-distribution detection for low-resolution radar micro-Doppler signatures}

\date{} 					

\author[1, 2]{\small Martin Bauw}
\author[1]{\small Santiago Velasco-Forero}
\author[1]{\small Jesus Angulo}
\author[2]{\small Claude Adnet}
\author[2]{\small Olivier Airiau}

\affil[1]{\footnotesize Center for Mathematical Morphology, Mines Paris, PSL University, France}
\affil[2]{\footnotesize Thales LAS France, Advanced Radar Concepts, Limours, France}

\renewcommand{\headeright}{HAL preprint}
\renewcommand{\undertitle}{HAL preprint}
\renewcommand{\shorttitle}{Near OODD for low-resolution radar micro-Doppler signatures}

\hypersetup{
pdftitle={Near out-of-distribution detection for low-resolution radar micro-Doppler signatures},
pdfauthor={Martin Bauw, Santiago Velasco-Forero, Jesus Angulo, Claude Adnet, Olivier Airiau},
pdfkeywords={Anomaly detection, out-of-distribution detection, micro-Doppler, radar target detection, deep learning},
}

\newcommand\norm[1]{\left\lVert#1\right\rVert}
\newcommand{\realset}{\mathbb{R}}

\begin{document}
\maketitle

\begin{abstract}
Near out-of-distribution detection (OODD) aims at discriminating semantically similar data points without the supervision required for classification. This paper puts forward an OODD use case for radar targets detection extensible to other kinds of sensors and detection scenarios. We emphasize the relevance of OODD and its specific supervision requirements for the detection of a multimodal, diverse targets class among other similar radar targets and clutter in real-life critical systems. We propose a comparison of deep and non-deep OODD methods on simulated low-resolution pulse radar micro-Doppler signatures, considering both a spectral and a covariance matrix input representation. The covariance representation aims at estimating whether dedicated second-order processing is appropriate to discriminate signatures. The potential contributions of labeled anomalies in training, self-supervised learning, contrastive learning insights and innovative training losses are discussed, and the impact of training set contamination caused by mislabelling is investigated.
\end{abstract}


{\let\thefootnote\relax\footnote{This preprint is an expanded version of a paper accepted at ECML PKDD 2022 and soon to appear in the conference proceedings edited by Springer Nature.}}

\section{Introduction}

Near out-of-distribution detection (OODD) aims at distinguishing one or several data classes from semantically similar data points. For instance, identifying samples from one class of CIFAR10 among samples of the other classes of the same dataset solves a near OODD task. On the other hand, separating CIFAR10 samples from MNIST samples is a far OODD task: there is no strong semantic proximity between the data points being separated. OODD defines a kind of anomaly detection (AD) since OODD can be seen as separating a normal class from infinitely diverse anomalies, with a training set only or mostly composed of normal samples, and anomalies being possibly semantically close to normal samples~\cite{ren2021simple}. This training paradigm relies on lower supervision requirements compared to supervised classification, for which each class calls for a representative set of samples in the training data. 

This work considers both unsupervised and semi-supervised AD (SAD). Unsupervised AD trains the model with a representative set of normal data samples, while semi-supervised AD also benefits from labeled anomalies~\cite{hendrycks2019oe,ruff2020deep} that can not be representative, since anomalies are by definition infinitely diverse. A distinction can however be observed between benefiting from far and near anomalies, in analogy with far and near OODD, to refine the discrimination training. The contribution of self-supervision will be taken into account through the supply of far artificial anomalies for additional supervision during training.

Near OODD constitutes an ideal mean to achieve radar targets discrimination, where an operator wants an alarm to be raised everytime specific targets of interest are detected. This implies discriminating between different kinds of planes, or ships, sometimes being quite similar from a radar perspective and without numerous labeled samples available. For example, two ships can have close hull and superstructure sizes, implying close radar cross-sections, even though their purpose and equipment on deck are completely different. Analogous observations could be made for helicopters, planes and drones. In an aerial radar context, whereas separating aerial vehicles would constitute a near OODD task, spotting weather-related clutter would define a far OODD. Such an OODD-based detection setup is directly applicable to other sensors.

The motivation behind the application of OODD methods to low-resolution pulse Doppler radar (PDR) signatures stems from the constraints of some air surveillance radars. Air surveillance PDRs with rotating antennas are required to produce very regular updates of the operational situation and to detect targets located at substantial ranges. The regular updates dictate the rotation rate and limit the number of pulses, and thus the number of Doppler spectrum bins, over which to integrate and refine a target characterization. The minimum effective range restricts the pulse repetition frequency (PRF), which in turns diminishes the range of velocities covered by the Doppler bins combined. The operating frequencies of air surveillance radars are such that they can not make up for this Doppler resolution loss~\cite{levanon2004radar}. This work aims at exploring the potential of machine learning to discriminate targets within these air surveillance radars limitations, using the targets Doppler spectrums. Refining radar targets discrimination with limited supervision is critical to enable the effective detection of targets usually hidden in cluttered domains, such as small and slow targets.

The AD methods examined will take a series of target Doppler spectrums as an input sample. This series is converted into a second-order representation through the computation of a covariance matrix to include an AD method adapted to process symmetric positive definite (SPD) inputs in our comparison. Radar Doppler signatures with sufficient resolution to reveal micro-Doppler spectrum modulations is a common way to achieve targets classification in the radar literature, notably when it comes to detecting drones hidden in clutter~\cite{brooks2018temporal,bjorklund2019target,gerard2021micro}. The processing of second-order representations is inspired by their recent use in the machine learning literature~\cite{huang2017riemannian,yu2017second}, including in radar processing~\cite{brooks2019hermitian,arnaudon2013riemannian}, and is part of the much larger and very active research on machine learning on Riemannian manifolds~\cite{brooks2019riemannian,bronstein2017geometric}.

This paper first details the simulation setup which generates the micro-Doppler dataset, then describes the OODD methods compared. Finally, an experimental section compares quantitatively various supervision scenarios involving SAD and self-supervision. The code for both the data generation and the OODD experiments is available\footnote{\url{https://github.com/Blupblupblup/Doppler-Signatures-Generation}\\ \url{https://github.com/Blupblupblup/Near-OOD-Doppler-Signatures}}. The code made available does not restrict itself to the experiments put forward in the current document, pieces of less successful experiments being kept for openness and in case they help the community experiment on the data with similar approaches.

\section{Micro-Doppler dataset}
\label{Dataset}

\begin{figure}
\includegraphics[width=0.85\textwidth]{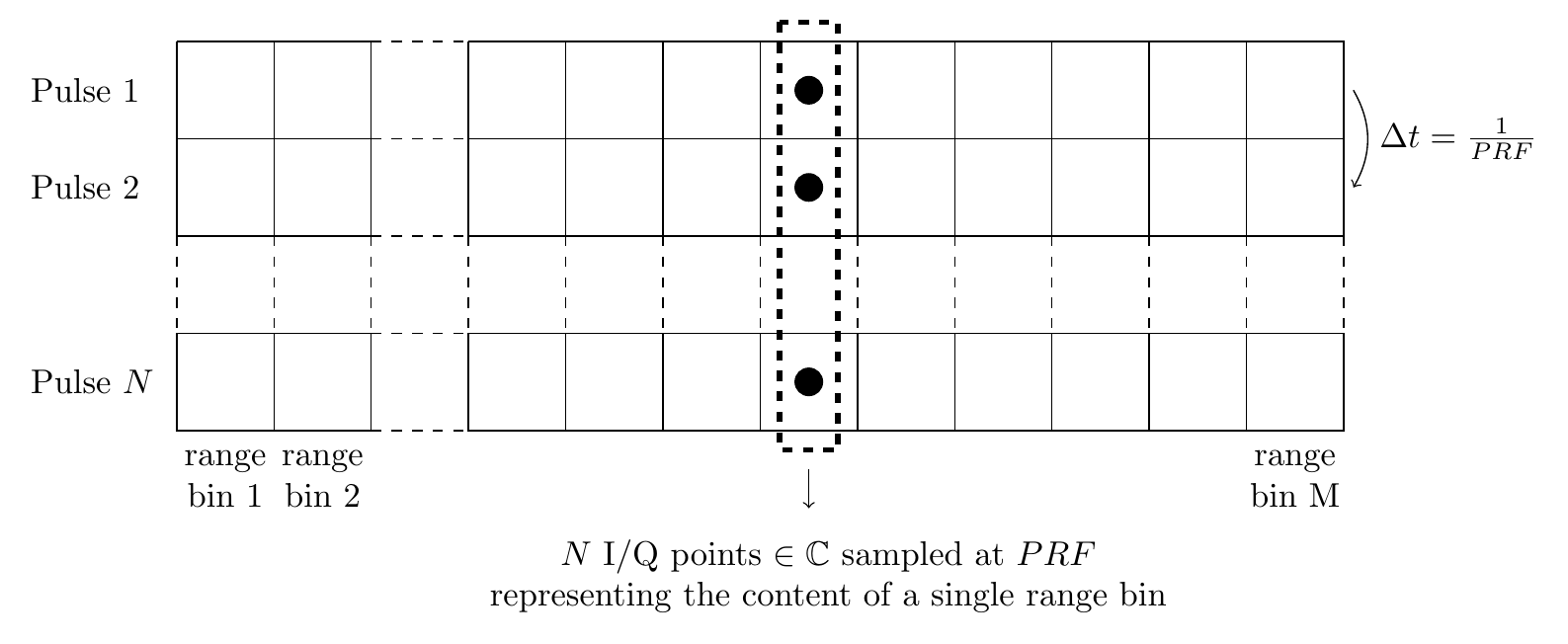}
\caption{Each pulse leads to one complex-valued I/Q sample per range bin, while each burst is composed of several pulses. Each range bin is thus associated with a complex-valued discrete signal with as many samples as there are pulses. Air surveillance radars with rotating antennas are required to provide regular situation updates in every direction, severely constraining the number of pulses per burst acceptable.}
\label{fig1}
\end{figure}

A PDR is a radar system that transmits bursts of modulated pulses, and after each pulse transmission waits for the pulse returns. The pulse returns are sampled and separated into range bins depending on the amount of time observed between transmission and reception. The spectral content of the sampled pulses is evaluated individually in each range bin, as depicted on Fig.~\ref{fig1}. This content translates into the Doppler information which amounts to a velocity descriptor: the mean Doppler shift reveals the target bulk speed, and the spectrum modulation its rotating blades. These Doppler features are available for each burst, under the assumption that the velocities detected in a given range bin change negligibly during a burst. The number of pulses in a burst, equating the number of samples available to compute a spectrum, determines the resolution of the Fourier bins or Doppler bins. The PRF sampling frequency defines the range of speeds covered by the spectrum. PDR signatures are generated by a MATLAB~\cite{matlab} simulation. The Doppler signatures are a series of periodograms, i.e. the evolution of spectral density over several bursts, one periodogram being computed per burst. Both periodograms and the series of periodograms can be called spectrum in the remainder of this paper. The samples on which the discrete Fourier transform is computed are sampled at the PRF frequency, i.e. one sample is available per pulse return for each range bin.

\begin{figure}
\includegraphics[width=0.6 \textwidth]{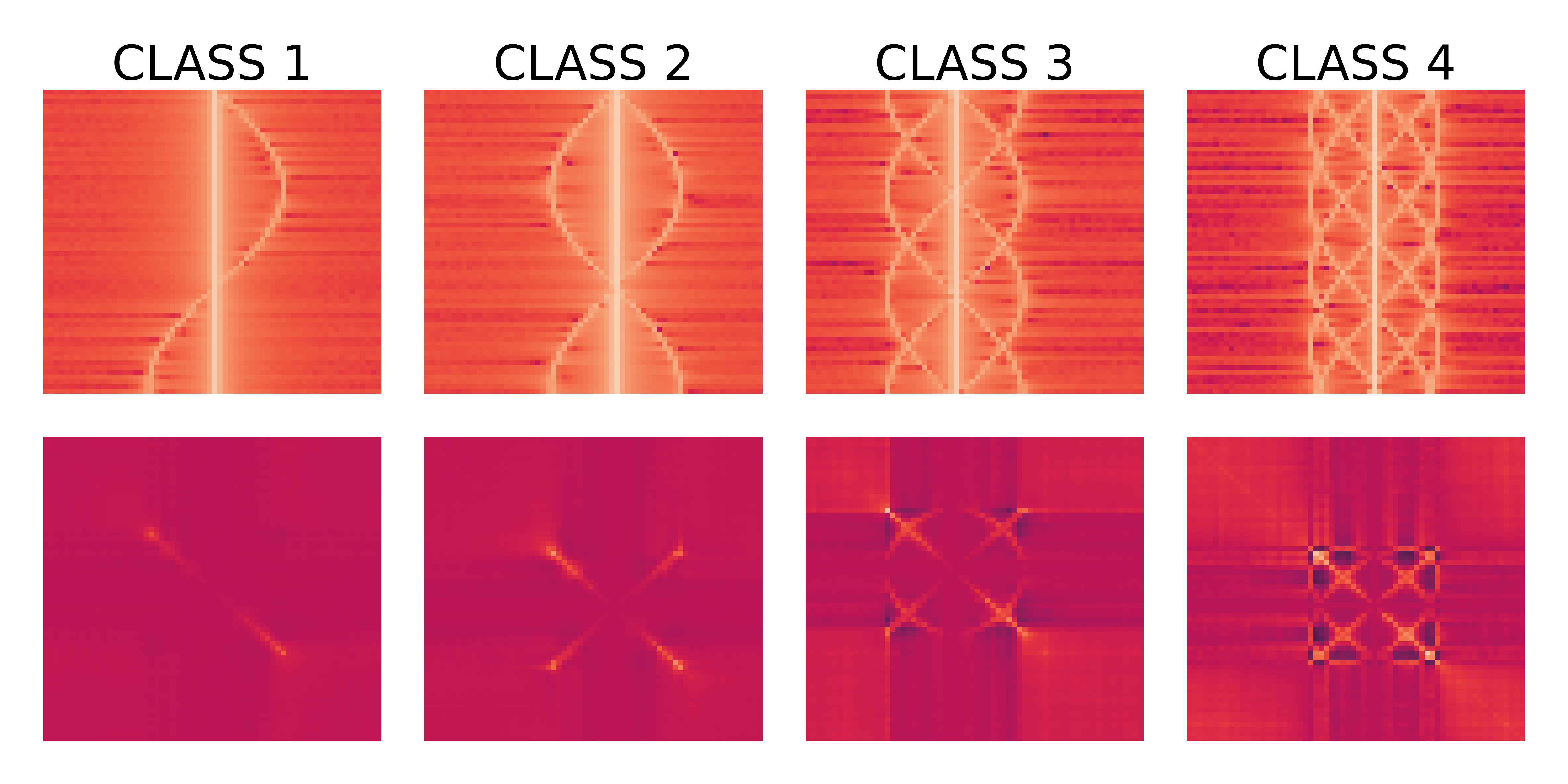}
\caption{One sample of each target class: the varying number of rotating blades defines the classes, the modulation pattern being easily singled out. The first line of images shows Doppler signatures, i.e. the time-varying periodogram of targets over 64 bursts of 64 pulses. On those images, each row is the periodogram computed over one burst, and each column a Fourier i.e. a Doppler bin. The second line contains the covariance SPD representation of the first line samples. The width of the Doppler modulations around the bulk speed on the periodograms varies within each class, as well as the bulk speed, the latter being portrayed by the central vertical illumination of the signature.}
\label{example-samples}
\end{figure}

The main parameters of the simulation are close to realistic radar and target characteristics. A carrier frequency of 5 GHz was selected, with a PRF of 50 KHz. An input sample is a Doppler signature extracted from 64 bursts of 64 pulses, i.e. 64 spectrums of 64 points, ensuring the full rank of the covariance matrix computed over non-normalized Doppler, i.e. Fourier, bins. The only simulation parameter changing across the classes of helicopter-like targets is the number of rotating blades: Doppler signatures are associated with either one, two, four or six rotating blades, as can be found on drones and radio-controlled helicopters. The quality of the dataset is visually verified: a non-expert human is easily able to distinguish the four target classes, confirming the discrimination task is feasible. The classes intrinsic diversity is ensured by receiver noise, blade size and revolutions-per-minute (RPM) respectively uniformly sampled in $\left[4.5,7\right]$ and $\left[450,650\right]$, and a bulk speed uniformly sampled so that the signature central frequency changes while staying approximately centered. The possible bulk speeds and rotor speeds are chosen in order for the main Doppler shift and the associated modulations to remain in the unambiguous speeds covered by the Doppler signatures~\cite{levanon2004radar}. Example signatures and their covariance representations are depicted for each class on Fig.~\ref{example-samples}. For each class, 3000 samples are simulated, thus creating a 12000-samples dataset. While small for the deep learning community, possessing thousands of relevant and labeled real radar detections would not be trivial in the radar industry, making larger simulated datasets less realistic for this use case.

\section{OODD methods}
\label{OOD}

This work compares deep and non-deep OODD methods, called shallow, including second-order methods harnessing the SPD representations provided by the covariance matrix of the signatures. The extension of the deep learning architectures discussed to SAD and self-supervised learning (SSL) is part of the comparison. The use of SSL here consists in the exploitation of a rotated version of every training signature belonging to the normal class in addition to its non-rotated version, whereas SAD amounts to the use of a small minority of actual anomalies taken in one of the other classes of the dataset. In the first case one creates artificial anomalous samples from the already available samples of a single normal class, whereas in the second case labeled anomalies stemming from real target classes are made available. To avoid confusion, one should note that this single normal class can be made of one or several target classes, which end up being considered as a unique normality. No SSL or SAD experiments were conducted on the SPD representations, since the SSL and SAD extensions of the deep methods are achieved through training loss modifications, and the SPD representations were confined to shallow baselines. In the previously described setup, SSL is nothing more than SAD with artificial data points provided by SSL transformations.

\subsection{Non-deep methods}
\label{non-deep-methods}

Common non-deep anomaly detection methods constitute our baselines: one-class support vector machines (OC-SVM)~\cite{scholkopf2001estimating}, isolation forests (IF)~\cite{liu2008isolation}, local outlier factor (LOF)~\cite{breunig2000lof} and random projections outlyingness (RPO)~\cite{donoho1992breakdown}. The three first methods are selected for their widespread use~\cite{chandola2009anomaly,ruff2018deep,bauw2020unsupervised}, and the diversity of the underlying algorithms. OC-SVM projects data points in a feature space where a hyperplane separates data points from the origin, thus creating a halfspace containing most samples. Samples whose representation lies outside of this halfspace are then considered to be anomalies. IF evaluates how easy it is to isolate data points in the feature space by recursively partitioning the representation space. The more partitions are required to isolate a data point, the more difficult it is to separate this point from other samples, and the less anomalous this point is. LOF uses the comparison of local densities in the feature space to determine whether a point is anomalous or not. Points that have local densities similar to the densities of their nearest neighbors are likely to be inliers, whereas an outlier will have a much different local density than its neighbors. RPO combines numerous normalized outlyingness measures over 1D projections with a $max$ estimator in order to produce a unique and robust multivariate outlyingness measure, which translates into the following quantity:
\begin{equation}
\label{RPO}
O(x;p, X) = \underset{u \in \mathbb{U}}{max} \dfrac{\vert u^Tx - MED(u^TX) \vert}{MAD(u^TX)}
\end{equation}
where $x$ is the data point we want to compute the outlyingness for, $p$ the number of random projections (RP) $u$ of unit norm gathered in the set $\mathbb{U}$, and $X$ the training data matrix. $MED$ stands for median, a location estimator, and $MAD$ for median absolute deviation, a spread estimator. The $max$ implies retaining only the worst outlyingness measure available among all the 1D projections, i.e. the worst normalized deviation from the projected median. In~\cite{velasco2012robust}, the asymptotic equivalence between Eq.~\ref{RPO} for a large number of RPs and the Mahalanobis distance (up to a constant factor) is established, with the motivation of obtaining an equivalent of the latter without computing a covariance matrix. This equivalence with the Mahalanobis distance indicates that RPO with enough RPs, after the $max$ integration over RPs, describes a normality ellipsoid in the input space, i.e. the representation space of $x$ (see Appendix~\ref{appendix-maha}). The RPO outlyingness actually leads to the definition of a statistical depth approximation~\cite{donoho1992breakdown,huber1985projection}, another quantity that orders data points from a given set from most to least normal.

\subsection{Deep methods}
\label{deep-methods}

The deep AD methods experimented on in this work are inspired by the deep support vector data description (SVDD) original paper~\cite{ruff2018deep}. Deep SVDD achieves one-class classification by concentrating latent space representations around a normality centroid with a neural network trained to minimize the distance of projected data samples to the centroid. The centroid is defined by the average of the initial forward pass of the training data, composed of normal samples. The intuition behind the use of Deep SVDD for AD is similar to the way one can detect anomalies with generative models: whereas generative models detect outliers because they are not as well reconstructed as normal samples, deep SVDD projects outliers further away from the normality centroid in the latent space. The AD score is thus the latent distance to the normality centroid after projection by the trained deep SVDD network. One can note that Deep SVDD is a deep learning adaptation of SVDD~\cite{tax2004support}, which can be equivalent to the OC-SVM method in our comparison if one uses a Gaussian kernel. The training loss of Deep SVDD for a sample of size $n$ with a neural network $\Phi$ with weights $W$ distributed over $L$ layers is as follows:

\begin{equation}
\label{DeepSVDD}
\min_{W} \Bigg[ \dfrac{1}{n} \sum_{i=1}^n \vert \vert \Phi(x_i;W) - c \vert \vert^2 + \dfrac{\lambda}{2} \sum_{l=1}^L \vert \vert W^l \vert \vert^2 \Bigg]
\end{equation}
where $c$ is the fixed normality centroid. The second term is a weights regularization controlled by $\lambda$. Deep SVDD naturally calls for a latent multisphere extension. An example of such an extension is Deep multi-sphere SVDD (MSVDD)~\cite{ghafoori2020deep}, which is part of our comparison. Deep MSVDD initializes numerous latent normality hyperspheres using k-means and progressively discards the irrelevant centroids during training. The relevance of latent hyperspheres is determined thanks to the cardinality of the latent cluster they encompass. The deep MSVDD training loss is:

\begin{equation}
\label{DeepMSVDD}
\underset{W,r_1 \dots r_K}{min} \Bigg[ \dfrac{1}{K} \sum_{k=1}^K r_k^2 + \dfrac{1}{\nu n} \sum_{i=1}^n max(0, \vert \vert \Phi(x_i;W) - c_j \vert \vert^2 - r_j^2) + \dfrac{\lambda}{2} \sum_{l=1}^L \vert \vert W^l \vert \vert^2 \Bigg]
\end{equation}

The first term minimizes the volume of hyperspheres of radius $r_k$, while the second is controlled by $\nu \in [0,1]$ and penalizes points lying outside of their assigned hypersphere, training samples being assigned to the nearest hypersphere of center $c_j$. A second Deep SVDD variant considered here is Deep RPO~\cite{bauw2021deep}, which replaces the latent Euclidean distance to the normality centroid with a RPs-based outlyingness measure adapted from Eq.~\ref{RPO} in the latent space, leading to the following loss:

\begin{equation}
\label{deepRPO}
\min_{W} 
\Bigg[ 
\dfrac{1}{n} \sum_{i=1}^n \left(\underset{u \in \mathbb{U}}{mean} \dfrac{\vert u^T \Phi(x_i;W) - MED(u^T\Phi(X;W)) \vert}{MAD(u^T\Phi(X;W))} \right)\\+ \frac{\lambda}{2} \sum_{l=1}^L \vert \vert W^l \vert \vert^2 
\Bigg]
\end{equation}

This training loss uses the outlyingness defined in Eq.~\ref{RPO} after the neural network encoding, with a $max$ estimator transformed into a $mean$ as suggested in~\cite{bauw2021deep} for better integration with the deep learning setup. The $mean$ estimator computes a mean over the set of RPs available to compute the latent outlyingness, while the $\frac{1}{n}$ computes a mean over the sample of size $n$. The use of a $mean$ instead of a $max$ removes the convergence to the Mahalanobis distance-inferred ellipsoid (up to a constant factor) already mentioned in the RPO definition for a large set of RPs. This RPO variant however remains affine invariant (see Appendix~\ref{appendix-invariance}). The loss nonetheless still combines 1D outlyingness measures individually centered by their median and normalized by their median absolute deviation, but with no ellipsoid-like score distribution guarantee in the input space once integrated (see Appendix~\ref{appendix-maha}). No square was applied to the first loss term, in accordance with the quantity put forward in~\cite{donoho1992breakdown}. Whereas Deep SVDD uses the Euclidean distance to the latent normality centroid as a test score, Deep MSVDD replaces this score with the Euclidean distance to the nearest latent centroid remaining after training, from which the associated radius is subtracted. Very often in our experiments, even with multimodal normality during training, only one latent sphere remains at the end of Deep MSVDD training. Deep RPO replaces the Euclidean distance score with an RPO computed in the encoding neural network output space.

SAD is achieved through outlier exposure~\cite{hendrycks2019oe,ruff2020deep}, which adds supervision to the training of the model thanks to the availability of few and non representative labeled anomalies. To take into account anomalies during training, Deep SAD~\cite{ruff2020deep} repels the outliers from the normality centroid by replacing the minimization of the distance to the centroid with the minimization of its inverse in the training loss. With $m$ labeled anomalies $\tilde{x}$ in a sample the Deep SVDD loss thus becomes, with a training objective balancing parameter $\eta$: 

\begin{equation}
\label{DeepSAD}
\min_{W} \Bigg[ \dfrac{1}{n+m} \sum_{i=1}^n \vert \vert \Phi(x_i;W) - c \vert \vert^2 \\ + \dfrac{\eta}{n+m} \sum_{j=1}^m (\vert \vert \Phi(\widetilde{x_j};W)-c\vert \vert^2)^{-1} +\dfrac{\lambda}{2} \sum_{l=1}^L \vert \vert W^l \vert \vert^2 \Bigg]
\end{equation}

Labeled anomalies in the training set need to be distinguished from potential unlabeled anomalies that are considered to be normal samples, which confuse the AD by contaminating the training set instead of providing supervision. This adaptation can be repeated for both Deep RPO and Deep MSVDD, although for Deep MSVDD the multiplicity of normality centers calls for an additional consideration on how to choose from which centroid the labeled anomalies should be repelled as long as several centroids are kept active. The experiments implementing Deep MSVDD adapted to SAD with an additional loss term for labeled anomalies were inconclusive, such an adaptation will therefore not be part of the presented results. The additional loss term either minimized the latent distance between anomalies and dedicated centroids, or maximized the latent distance between anomalies and normality centroids. Once adapted to SAD, the Deep RPO loss becomes, similarly to the transformation that led to Eq.~\ref{DeepSAD}:

\begin{equation}
\label{deepRPO-SAD}
\begin{split}
\min_{W} 
\Bigg[ 
\dfrac{1}{n+m} \sum_{i=1}^n \left(\underset{u \in \mathbb{U}}{mean} \dfrac{\vert u^T \Phi(x_i;W) - MED(u^T\Phi(X;W)) \vert}{MAD(u^T\Phi(X;W))} \right)\\+ 
\dfrac{\eta}{n+m} \sum_{j=1}^m \left(\underset{u \in \mathbb{U}}{mean} \dfrac{\vert u^T \Phi(\widetilde{x_j};W) - MED(u^T\Phi(X;W)) \vert}{MAD(u^T\Phi(X;W))} \right)^{-1}\\+
\frac{\lambda}{2} \sum_{l=1}^L \vert \vert W^l \vert \vert^2 
\Bigg]
\end{split}
\end{equation} 

Arbitrary sets of outliers could not be completely gathered around a reference point since they are not necessarily concentrated in a common mode~\cite{ruff2020deep,steinwart2005classification}. However, this does not forbid the concentration of identified modes among labeled anomalies close to dedicated centroids to provide additional supervision during training, a case which is part of the experiments presented. The possibly arbitrary distribution of normal and anomalous centroids and the relative distance between the centroids adds a way to use prior information regarding the proximity between the training samples. Such a setup can seem close to classification with rejection~\cite{hendrycks2019oe,bartlett2008classification}, since the concentration of data points around dedicated normal and anomalous centroids can be interpreted as classification while the data points attached to no centroid and thus supposedly repelled from all centroids by the trained network constitutes a rejection. This parallel with classification with rejection is not necessarily relevant since the availability of labeled anomalies to train AD methods is usually very limited if not nonexistent. In contrast, supervised classification of identified data modes would imply rich, representative and relatively balanced datasets for each latent mode. The limited availability of labeled anomalies applies to actual anomalies and not to artificial anomalies provided by the transformation of existing training samples i.e. through self-supervision. With proper transformations self-supervision can produce as many labeled anomalies for training as there are normal samples, or even more if each normal sample is transformed multiple times. However this does not overcome the lack of representativeness of labeled anomalies. This is also made difficult since the choice of transformations requires expert knowledge.

The reunion of normal latent representations achieved through the deep one-class classification methods mentioned is analogous to the alignment principle put forward in~\cite{wang2020understanding}, which also argued for a latent uniformity. Whereas the alignment principle compels similar samples to be assigned similar representations, the uniformity principle demands the preservation of maximal information. One way to achieve that according to~\cite{wang2020understanding} is to push all features away from each other on the unit hypersphere to intuitively facilitate a uniform distribution. The extension of the Deep SVDD loss to encourage a form of latent uniformity using the pairwise distance between normal samples during training was investigated without ever improving the baselines. The experiments conducted to evaluate the contribution of a pairwise distance of normal samples latent representations loss term revolved around the following training loss format, where the term tasked with enforcing latent uniformity is weighted using $\beta$ and was expected to be judiciously balanced with the overall latent concentration:

\begin{equation}
\label{DeepSVDD-uniformity}
\min_{W} 
\Bigg[ \dfrac{1}{n} \sum_i^n \left\Vert \Phi(x_i;W) - c \right\Vert^2 + \dfrac{\beta}{n} \sum_{i \ne j}^n \left( \left\Vert x_i - x_j \right\Vert^2 \right)^{-1} + \dfrac{\lambda}{2} \sum_{l=1}^L \vert \vert W^l \vert \vert^2 \Bigg]
\end{equation}

The failure to make a loss term enforce a form of latent uniformity could signal the necessity of associating such a constraint with latent representations confined to a relevant manifold.

\subsection{Riemannian methods for covariance matrices}
\label{riemannian-methods}

Two SPD-specific AD approaches were considered. The first approach consists in replacing the principal component analysis (PCA) dimensionality reduction preceding shallow AD with an SPD manifold-aware tangent PCA (tPCA). The tPCA projects SPD points on the tangent space of the Fréchet mean, a Riemannian mean which allows to compute an SPD mean, keeping the computed centroid on the Riemannian manifold naturally occupied by the data. Using tPCA offers the advantage of being sensible to the manifold on which the input samples lie, but implies that input data is centered around the Riemannian mean and not too scattered. This makes tPCA a questionable choice when the objective set is AD with multimodal normality~\cite{pennec2018barycentric}, something that is part of the experiments put forward in this work. In other words, although the linear approximation of the SPD inputs around its Riemannian mean provides us with a manifold-aware dimensionality reduction, the distribution of the inputs in the SPD matrices manifold may lead this approximation to be excessively inaccurate, leading to irrelevant reduced representations. This is due to the linear approximation not preserving the Riemannian distances between points~\cite{sommer2010manifold}. Nonetheless, the Euclidean PCA being a common tool in the shallow AD literature, tPCA remains a relevant candidate for this study since it enables us to take a step back with respect to non-deep dimensionality reduction.

The second SPD-specific approach defines a Riemannian equivalent to Deep SVDD: inspired by recent work on SPD neural networks~\cite{lou2020differentiating,brooks2019riemannian,huang2017riemannian,yu2017second}, which learn intermediate representations while keeping them on the SPD matrices manifold, a Deep SVDD SPD would transform input covariance matrices and project the latter into a latent space comprised within the SPD manifold. Taking into account SAD and SSL labeled anomalies during training was expected to be done as for the semi and self-supervised adaptations of Deep SVDD described earlier, where labeled anomalies are pushed away from the latent normality centroid thanks to an inverse distance term in the loss. For Deep SVDD SPD, the distance would be a geometry-aware distance such as the Log-Euclidean distance. Despite diverse attempts to make such a Deep SVDD SPD model work, with and without geometry-aware non-linearities in the neural network architecture, no effective learning was achieved on our dataset. This second approach will therefore be missing from the reported experimental results. Since this approach defined the ReEig~\cite{huang2017riemannian} non-linearity rectifying small eigenvalues of SPD representations, the related shallow AD approach using the norm of the last PCA components as an anomaly score was also considered. This \emph{negated PCA} is motivated by the possibility that, in one-class classification where fitting occurs on normal data only, the first principal components responsible for most of the variance in normal data are not the most discriminating ones when it comes to distinguishing normal samples from anomalies~\cite{miolane2021iclr,rippel2021modeling}. This approach was applied to both spectral and covariance representations, with the PCA and tPCA last components respectively, but was discarded as well due to poor performances. The latter indicate that anomalous samples are close enough to the normal ones for their information to be carried in similar components, emphasizing the near OODD nature of the discrimination pursued.

\section{Experiments}
\label{Experiments}

AD experiments are conducted for two setups: a first setup where normality is made of one target class, and a second setup where normality is made of two target classes. When a bimodal normality is experimented on, the normal classes are balanced. Moreover, the number of normal modes is not given in any way to the AD methods, making the experiments closer to the arbitrary and, to a certain extent, unspecified one-class classification useful to a radar operator. Within the simulated dataset, 90\% of the samples are used to create the training set, while the rest is equally divided to create the validation and test sets. All non-deep AD methods are applied after a preliminary dimensionality reduction, which is either PCA or tPCA. The number of RPs used to compute the outlyingness score with RPO and Deep RPO is the same and set to 1000, even though the estimator used differs between the shallow and the deep approach. All deep experiments were run on a single GPU, which was either a NVIDIA Tesla P100 or a NVIDIA RTX 2080. In both cases running one deep AD setup for 10 seeds took approximately one hour. Non-deep experiments were CPU-intensive and also required around one hour for 10 seeds on a high-end multi-core CPU.

\begin{figure}
\floatbox[{\capbeside\thisfloatsetup{capbesideposition={right,top},capbesidewidth=6cm}}]{figure}[\FBwidth]
{\caption{Random samples of the fourth class after the preprocessing erasing the irrelevant background, which makes the dataset closer to the MNIST data format. One can notice the varying modulation width of the target spectrum and central Doppler shift. The fourth class has the highest number of rotating blades on the helicopter-like target, hence the higher complexity of the pattern. These samples illustrate the input format of the various AD methods compared in this work, and are min-max normalized so that their values belong to $[0,1]$. Only the inputs of the Riemannian setup where shallow learning AD is used on SPD inputs after tPCA uses a different input format, where the input covariance is not normalized.}\label{preprocessed-samples}}
{\includegraphics[width=5.5cm]{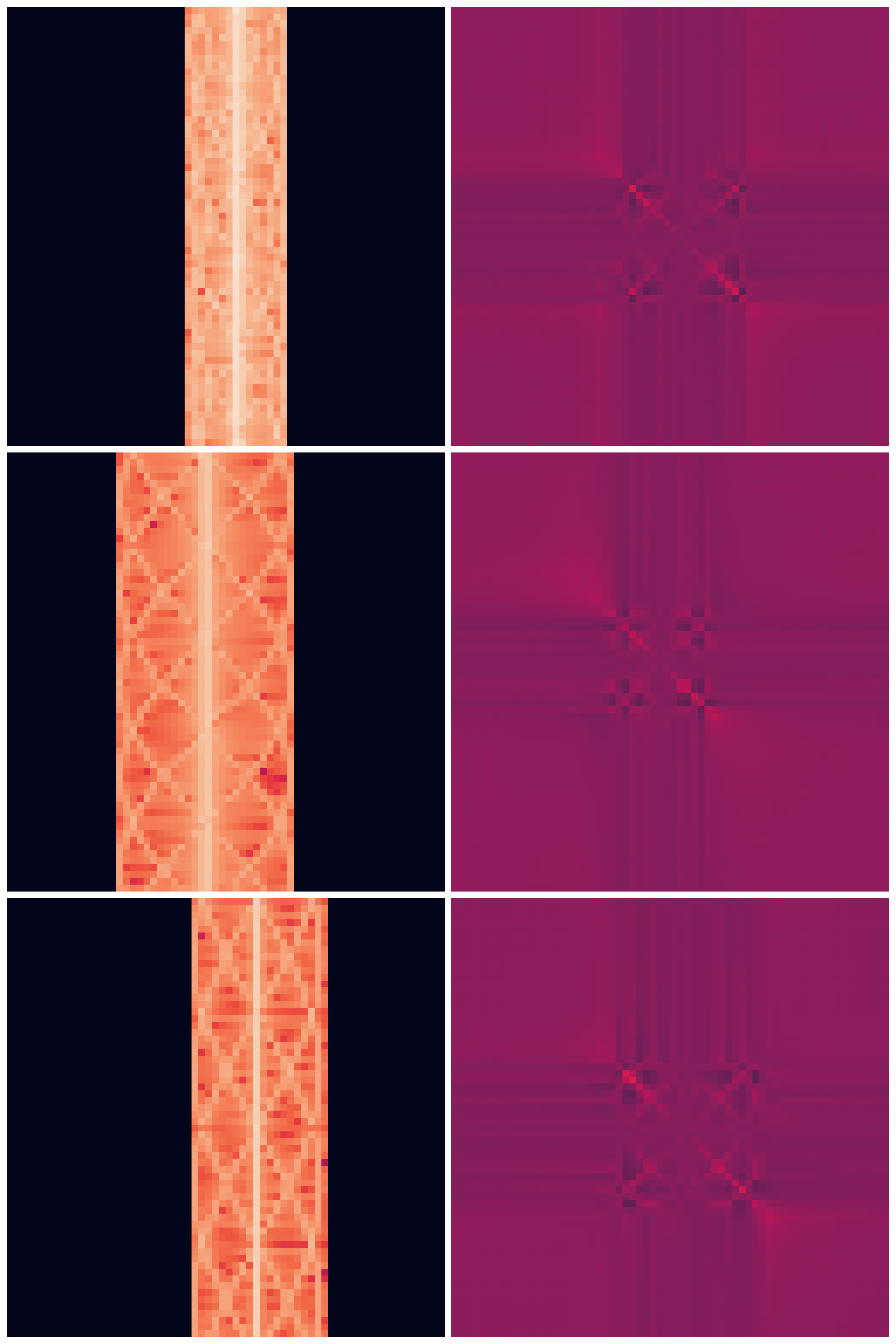}}
\end{figure}

\paragraph{Preprocessing} This work is inspired by~\cite{ruff2018deep}, which experimented on MNIST, a dataset in which samples are images of objects without background or irrelevant patterns. In order to guarantee a relevant neural architecture choice, this kind of input format is deliberately reproduced. The idea of creating MNIST-like benchmarks has been of interest in different scientific communities such as biomedical images~\cite{yang2021medmnist} and 3D modeling~\cite{jimenez2016unsupervised}. The series of periodograms, i.e. non-SPD representations are therefore preprocessed such that only the columns with top 15\% values in them are kept, this operation being done after a switch to logarithmic scale. This results in periodograms where only the active Doppler bins, portraying target bulk speed and micro-Doppler modulations, have non-zero value. Only a grayscale region of interest (ROI) remains in the input matrix with various Doppler shifts and modulation widths, examples of which are shown on Figure~\ref{preprocessed-samples}. This preprocessing leads to the "(SP)" input format as indicated in the results tables, and is complementary to the covariance representation. Covariance matrices are computed without such preprocessing, except for the switch to logarithmic scale which precedes the covariance computation. Comparing covariance-based OODD to OODD on spectral representations is fair since both representations stem from the same inputs, the covariance only implying an additional transformation of the input before training the AD. All input data is min-max normalized except for the covariance matrices used by tPCA.

\paragraph{Deep learning experiments} The test AUC score of the best validation epoch in terms of AUC is retained, in line with~\cite{chong2020simple}. All experiments were conducted with large $1000$ samples batches, which stabilizes the evolution of the train, validation and test AUCs during training. The training is conducted during 300 epochs, the last 100 epochs being fine-tuning epochs with a reduced learning rate, a setup close to the one in~\cite{ruff2018deep}. As was suggested in~\cite{chong2020simple} a relatively small learning rate of $10^{-4}$ is chosen to help avoid the latent normality hypersphere collapse, i.e. the convergence to a constant projection point in the latent space in the non-SAD and non-SSL cases, with $\lambda=10^{-6}$. Such a latent normality collapse is made impossible when SAD or SSL samples are concentrated around dedicated centroids or scattered away from normality centroids, since the network is then trained to disperse representations. Loss terms integrating labeled anomalies for extra training supervision are balanced with $\eta=1$, and for Deep MSVDD $\nu=0.1$. Hyperparameters are kept constant across all experiments conducted, in order to ensure fair comparisons. In the results tables, the second and third columns indicate whether SAD and SSL samples were used for additional supervision during training, and describe how such samples affected the training loss if present. When the SAD or SSL loss term is defined by a centroid, it means that the distance to the mentioned centroid is minimized during training, whereas "away" implies the projection of the SAD or SSL samples are repelled from the normality centroid thanks to an inverse distance as described previously. For example, the first line of the second part of Table~\ref{sad-ssl-results} describes an experiment where SAD samples are concentrated around the SAD samples latent centroid, and SSL samples concentrated around the SSL samples latent centroid. Centroids are computed, as for the normal training samples, with the averaging of an initial forward pass, therefore yielding the average latent representation. 

\paragraph{Non-deep learning experiments} Shallow AD conducted on the covariance representation after a common PCA uses the upper triangular part of the min-max normalized input as a starting point, avoiding redundant values. This contrasts with the Riemannian approach replacing PCA with the tPCA, the latter requiring the raw SPD representation. Furthermore, shallow approaches were also tried on the periodograms individually, where each row of an input signature, i.e. one vector of Doppler bins described for one burst, was given a score, the complete signature being then given the mean score of all its periodograms. This ensemble method did not yield relevant results and is therefore missing from our comparison. Such an approach ignores the order of periodograms in signatures.

\paragraph{Neural network architecture} While the MNIST-like input format is thus replicated, the 2D features remain specific to radar signal processing and may therefore benefit from a different neural network architecture. Several neural networks architectures were considered, including architectures beginning with wider square and rectangular convolutions extended along the (vertical) bursts input axis, with none of the investigated architectures scoring systematically higher than the Fashion-MNIST architecture from the original Deep SAD work~\cite{ruff2020deep}, which was only modified in order to handle the larger input size. The latter was consequently selected to produce the presented results. This architecture projects data with two convolutional layers followed by two dense layers, each layer being separated from the next one by a batch normalization and a leaky ReLU activation. The outputs of the two convolutional layers are additionally passed through a 2D max-pooling layer.

\paragraph{Riemannian AD} The tPCA was computed thanks to the dedicated Geomstats\footnote{\url{https://geomstats.github.io/}}~\cite{geomstats} function, while experiments implementing a Riemannian equivalent of Deep SVDD were conducted using the SPD neural networks library torchspdnet\footnote{\url{https://gitlab.lip6.fr/schwander/torchspdnet}}~\cite{brooks2019second}. The AD experiments based on a SPD neural network ending up inconclusive, they are not part of the results tables.

\subsection{Unsupervised OODD with shallow and deep learning}
\label{unsupervised-AD}

Unsupervised AD results, for which the training is only supervised by normal training samples, are presented in Table~\ref{unsupervised-results}. These results indicate the superiority of deep learning for the OODD task considered, while demonstrating the substantial contribution of geometry-aware dimensionality reduction through the use of tPCA for non-deep AD. RPO is kept in Table~\ref{unsupervised-results} even though it does not achieve useful discrimination because it is the shallow equivalent of Deep RPO, one of the highlighted deep AD methods, deprived of the neural network encoder and with a $max$ estimator instead of a $mean$, as was previously justified. Deep MSVDD does not lead to the best performances, and is as effective as Deep SVDD and Deep RPO, which could seem surprising at least when normality is made of two target classes. Indeed, since Deep MSVDD has the possibility to use several disjointed hyperspheres to capture the latent normality distribution, one could expect it to better model more complex, e.g. multimodal, normality. 

\begin{table}
\caption{Unsupervised AD experiments results (average test AUCs in \% $\pm$ StdDevs over ten seeds). These machine learning methods are trained on fully normal training sets, without labeled anomalies for SAD or self-supervision transformations. The four last methods are our deep AD baselines, trained on normalized spectral representations only. Deep MSVDD "mean best" indicates the neural network was trained using a simpler loss, analogous to the Deep SVDD loss, where only the distance to the best latent normality centroid is minimized, thus discarding the radius loss term. PCA and tPCA indicate that the AD model is trained after an initial dimensionality reduction, which is either PCA or tangent PCA.}
\label{unsupervised-results}
\begin{tabular}{|l|c|c|c|c|c|c|}
\hline
AD method (input format) & SAD loss & SSL loss & Mean test AUC (1 mode) & Mean test AUC (2 modes) \\
\hline

OC-SVM (SP-PCA)   & / & / & 49.16 $\pm$ 26.69 & 45.48 $\pm$ 27.53 \\
OC-SVM (SPD-PCA)  & / & / & 64.68 $\pm$ 9.10 & 58.23 $\pm$ 15.12 \\
OC-SVM (SPD-tPCA) & / & / & 57.59 $\pm$ 3.91 & 55.33 $\pm$ 9.48\\

IF (SP-PCA)   & / & / & 50.96 $\pm$ 17.37 & 48.50 $\pm$ 18.76 \\
IF (SPD-PCA)  & / & / & 52.36 $\pm$ 22.47 & 47.50 $\pm$ 20.32 \\
IF (SPD-tPCA) & / & / & 66.91 $\pm$ 9.65 & 61.23 $\pm$ 12.65 \\

LOF (SP-PCA)   & / & / & 56.80 $\pm$ 2.38 & 61.55 $\pm$ 10.29 \\
LOF (SPD-PCA)  & / & / & 66.44 $\pm$ 21.37 & 65.83 $\pm$ 19.52 \\
LOF (SPD-tPCA) & / & / & 78.38 $\pm$ 8.86 & 73.56 $\pm$ 10.09 \\

RPO (SP-PCA)   & / & / & 49.61 $\pm$ 6.89 & 50.43 $\pm$ 7.13 \\
RPO (SPD-PCA)  & / & / & 51.08 $\pm$ 19.66 & 54.95 $\pm$ 17.58 \\
RPO (SPD-tPCA) & / & / & 33.97 $\pm$ 7.36 & 38.08 $\pm$ 14.58 \\

\hline
Deep SVDD (SP)              & no SAD & no SSL & 83.03 $\pm$ 6.83 & \textbf{78.29} $\pm$ \textbf{6.68} \\
Deep MSVDD (SP)             & no SAD & no SSL & 82.27 $\pm$ 9.67 & \textbf{78.30} $\pm$ \textbf{8.28} \\
Deep MSVDD "mean best" (SP) & no SAD & no SSL & 82.29 $\pm$ 7.20 & 78.02 $\pm$ 6.80 \\
Deep RPO (SP)               & no SAD & no SSL & \textbf{83.60} $\pm$ \textbf{5.35} & 78.13 $\pm$ 6.02\\
\hline

\end{tabular}
\end{table}

\subsection{Potential contribution of SAD and SSL}
\label{SAD-SSL-potential}

The contribution of additional supervision during training through the introduction of SAD samples and SSL samples is examined in Table~\ref{sad-ssl-results}. Regarding SAD experiments, labeled anomalies will be taken from a single anomalous class for simplicity, and because only four classes are being separated, this avoids unrealistic experiments where labeled anomalies from every anomalous class are seen during training. When SAD samples are used during training, labeled anomalies represent one percent of the original training set size. This respects the spirit of SAD, for which labeled anomalies can only be a minority of training samples, which is not representative of anomalies. This is especially realistic in the radar processing setup initially described where labeled detections would rarely be available. SSL samples are generated thanks to a rotation of the spectral input format, rendering the latter absurd but encouraging better features extraction since the network is asked to separate similar patterns with different orientations. SSL samples are as numerous as normal training samples, implying they don't define a minority of labeled anomalies for training as SAD samples do, when they are taken into account.

Individually, SAD samples lead to better performances than SSL ones, but the best results are obtained when combining the two sets of samples for maximal training supervision. Deep SVDD appears to be substantially better at taking advantage of the additional supervision provided by SAD and SSL samples. Quite surprisingly for a radar operator, the best test AUC is obtained when SSL samples are concentrated around a specialized centroid while SAD samples are repelled from the normality centroid. Indeed, SSL samples being the only absurd samples considered in our experiments radarwise, it could seem more intuitive to project SAD samples, which remain valid targets, next to a dedicated centroid while repelling SSL samples. Likewise, on an ideal outlyingness scale, SSL samples should be further away from normality than SAD samples. This counter-intuitive performance could stem from the test set which only evaluates the separation of targets in a near OODD context. No invalid target representation, like the SSL samples are, is present in the test set, only valid representation from the four target classes make up the latter. This is consistent with the application put forward in this study: use OODD to discriminate between various kinds of radar detections.

\begin{table}
\caption{Experiments with additional supervision provided by SAD and/or SSL labeled samples during training (average test AUCs in \% $\pm$ StdDevs over ten seeds). When available, SAD samples are the equivalent of one percent of the normal training samples in quantity. The first half of the Table reports performances where only one of the two kinds of additional supervision is leveraged, while the second half describes the performances for setups where both SAD and SSL labeled samples contribute to the model training. Each couple of lines compares Deep SVDD and Deep RPO in a shared AD supervision setup, thus allowing a direct comparison. \textit{c.} stands for centroid.}
\label{sad-ssl-results}
\begin{tabular}{|l|l|l|c|c|c|c|}
\hline
AD method (input format) & SAD loss & SSL loss & Mean test AUC (1 mode) & Mean test AUC (2 modes) \\
\hline
Deep SVDD (SP) & no SAD & SSL c.       & 86.79 $\pm$ 6.54 & 83.91 $\pm$ 7.92 \\
Deep RPO (SP)  & no SAD & SSL c.       & 88.70 $\pm$ 5.10 & 84.59 $\pm$ 8.54 \\
\hline
Deep SVDD (SP) & no SAD & away               & 81.43 $\pm$ 8.62 & 77.01 $\pm$ 8.20 \\
Deep RPO (SP)  & no SAD & away               & 80.21 $\pm$ 9.06 & 78.93 $\pm$ 9.39 \\
\hline
Deep SVDD (SP) & SAD c.                & no SSL & 86.79 $\pm$ 8.94 & 87.65 $\pm$ 6.44 \\
Deep RPO (SP)  & SAD c.                & no SSL & 81.38 $\pm$ 6.09 & 76.45 $\pm$ 6.30 \\
\hline
Deep SVDD (SP) & away & no SSL               & 93.93 $\pm$ 4.82 & 93.50 $\pm$ 7.61 \\
Deep RPO (SP)  & away & no SSL               & 84.19 $\pm$ 5.32 & 80.37 $\pm$ 7.22 \\
\hline
\hline
Deep SVDD (SP) & SAD c. & SSL c. & 91.00 $\pm$ 6.45 & 90.51 $\pm$ 7.38 \\
Deep RPO (SP)  & SAD c. & SSL c. & 87.79 $\pm$ 5.81 & 82.69 $\pm$ 8.51 \\
\hline
Deep SVDD (SP) & SAD c. & away         & 89.98 $\pm$ 7.79 & 91.03 $\pm$ 6.71 \\
Deep RPO (SP)  & SAD c. & away         & 78.86 $\pm$ 9.10 & 79.11 $\pm$ 9.64 \\
\hline
Deep SVDD (SP) & away & SSL c.         & \textbf{95.06} $\pm$ \textbf{4.20} & 93.91 $\pm$ 7.31 \\
Deep RPO (SP)  & away & SSL c.         & 89.82 $\pm$ 5.21 & 87.17 $\pm$ 8.17 \\
\hline
Deep SVDD (SP) & away & away                 & 94.63 $\pm$ 4.31 & \textbf{94.02} $\pm$ \textbf{7.30} \\
Deep RPO (SP)  & away & away                 & 90.91 $\pm$ 5.94 & 92.69 $\pm$ 7.98 \\
\hline

\end{tabular}
\end{table}


\begin{figure}
\includegraphics[width=17cm]{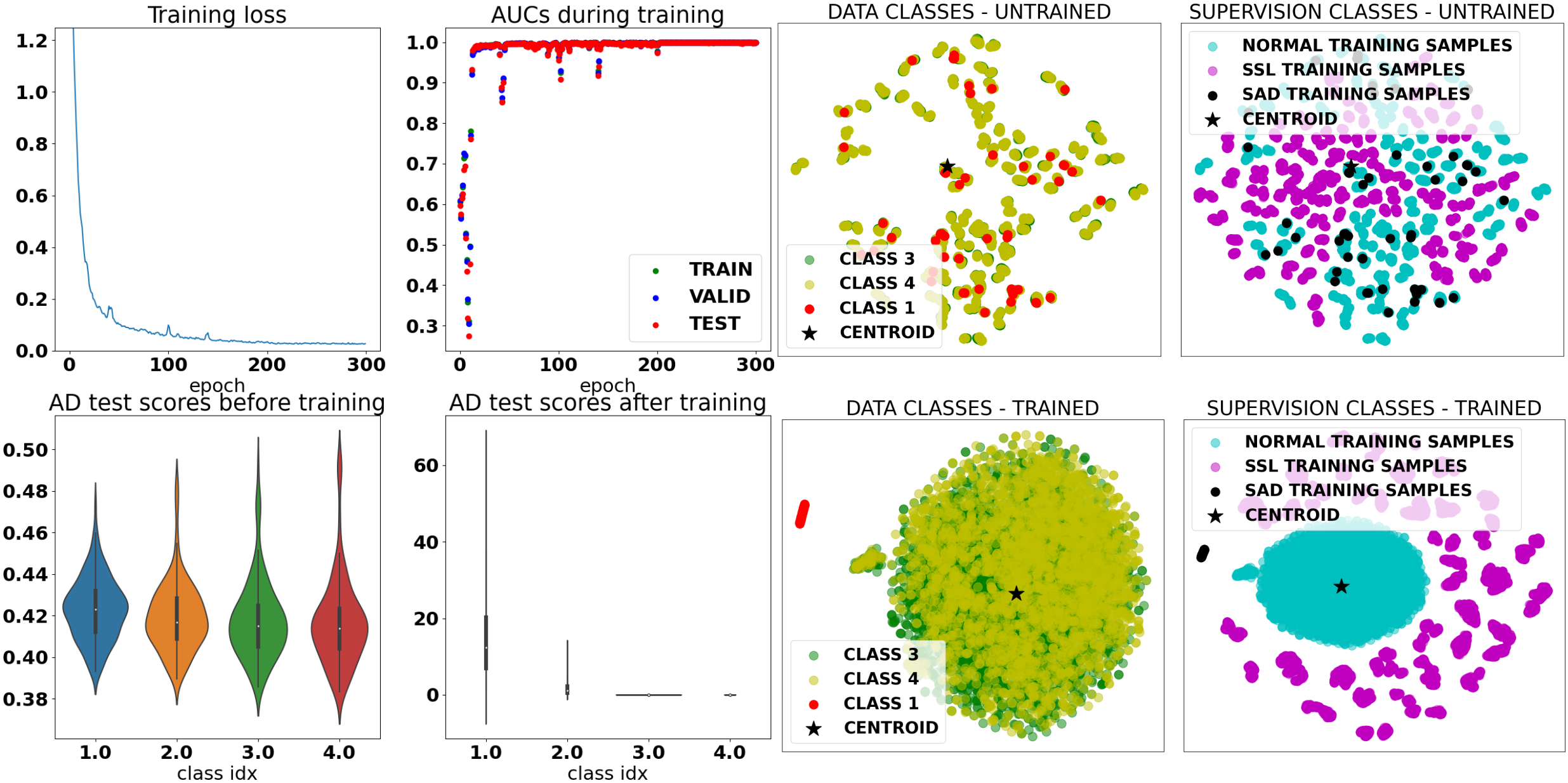}
\caption{\textbf{Left} - Training metrics of a successful run where normal samples are concentrated around their average initial projection, and SAD and SSL samples are pushed away thanks to a loss term using the inverse of the distance with respect to the normality latent centroid. This is one of the most successful setups in Table~\ref{sad-ssl-results}, and one of the easiest AD experiments since the two classes defining normality here are class 3 (four blades are responsible for the modulation pattern around the central Doppler shift) and class 4 (six blades are responsible for the modulation pattern around the central Doppler shift), meaning the separation with the other classes deemed anomalous is actually a binary modulation complexity threshold. One of the contributions of the SAD and SSL supervisions can be observed on the evolution of AUCs during training: no AUC collapse can be seen during training, discarding the possibility of a latent distribution collapse during training. Experiments showed that large training batches contributed to stable AUCs growth. Spikes in the training loss match the drops in AUCs. \textbf{Right} - Latent distribution of the training samples visualized in 2D using t-SNE after projection by the untrained (top) and the trained neural network (bottom). One can notice that normal training samples from both normal classes are completely mixed up with the minority of SAD labeled anomalies from class 1 in red (one blade), semantically similar, whereas SSL samples which are rotated normal training samples are already gathered in their own latent subclusters. SAD labeled anomalies end up well separated after training.}
\label{training-metrics}
\end{figure}


\subsection{Training with a contaminated training set}
\label{training-contaminated}

Unsupervised AD refers to the experiments of Table~\ref{unsupervised-results} where only training samples assumed to be normal supervise the training of the neural network. Real-life datasets, labeled by algorithms or experts, are unlikely to respect that assumption and will suffer from contamination of normal samples with unlabeled anomalies. The results in Table~\ref{contamination-experiments} depict how sensible the deep AD methods previously introduced are to training set contamination. The contamination is carried out using the one percent SAD samples already used for SAD experiments. While in the SAD experiments SAD samples were repelled from the normality centroid or concentrated next to their dedicated latent reference point, here they will be processed as normal samples. SSL samples again appear to better contribute to improving AD when concentrated next to a specialized centroid, while the performance drop due to contamination does not seem to be particularly stronger for one of the approaches considered.

\begin{table}
\caption{Contamination experiments results (average test AUCs in \% $\pm$ StdDevs over ten seeds): the SAD labeled anomalies are integrated within the training samples and taken into account as normal samples during training, thus no SAD loss term is used for SAD samples. The contamination rate is one percent, i.e. the equivalent of one percent of the normal training samples in labeled anomalies is added to confuse the AD. \textit{c.} stands for centroid.}
\label{contamination-experiments}
\begin{tabular}{|l|c|c|c|c|c|c|}
\hline
AD method (input format) & SAD loss & SSL loss & Mean test AUC (1 mode) & Mean test AUC (2 modes) \\
\hline
Deep SVDD (SP)              & no SAD & no SSL       & 80.76 $\pm$ 7.11 & 76.02 $\pm$ 6.66 \\
Deep MSVDD (SP)             & no SAD & no SSL       & 78.31 $\pm$ 11.18 & 74.49 $\pm$ 9.13 \\
Deep MSVDD "mean best" (SP) & no SAD & no SSL       & 79.84 $\pm$ 7.82 & 74.89 $\pm$ 7.01 \\
Deep RPO (SP)               & no SAD & no SSL       & 81.29 $\pm$ 5.92 & 74.82 $\pm$ 5.89 \\
\hline
Deep SVDD (SP) & no SAD & SSL c. & 85.34 $\pm$ 6.85 & 81.36 $\pm$ 7.47 \\
Deep RPO (SP)  & no SAD & SSL c. & \textbf{86.66} $\pm$ \textbf{6.41} & \textbf{82.78} $\pm$ \textbf{8.25} \\
\hline
Deep SVDD (SP) & no SAD & away         & 79.62 $\pm$ 9.02 & 75.38 $\pm$ 8.28 \\
Deep RPO (SP)  & no SAD & away         & 76.16 $\pm$ 9.87 & 76.56 $\pm$ 8.69 \\
\hline

\end{tabular}
\end{table}

\section{Conclusion}


The near OODD performances of various deep and non-deep, unsupervised and semi-supervised AD methods were compared on a radar Doppler signatures simulated dataset. Deep AD approaches were evaluated in various supervision setups, which revealed the relevance of combining a minority of labeled anomalies with transformed normal training samples to improve semi-supervised near OODD performances, and avoid latent normality distribution collapse. Among the limitations of our study, one can note the lack of OODD experiments on a multimodal normal training set with unbalanced normal classes, which would make the OODD task more realistic. Additionally, further reducing the Doppler resolution and the number of training samples available could make the targets discrimination task closer to reality. Taking into account a varying Doppler resolution, \emph{i.e.} a varying number of pulses per burst, could also improve the relevance of the proposed discrimination. The benefits of deep learning clearly showed, and while not leading to the best overall performances, geometry-aware processing with tangent PCA proved to be the source of a substantial improvement for non-deep AD.

\subsubsection*{Acknowledgements} This work was supported by the French Defense Innovation Agency (Cifre-Défense 001/2019/AID). The authors additionally thank Maxime Bombar and Lev-Arcady Sellem for their constructive comments regarding the mathematical properties of RPO-MEAN.

\bibliographystyle{plain}
\bibliography{references}  

\appendix

\section*{Appendix}

\section{About the relation between RPO, Deep RPO and the Mahalanobis distance}
\label{appendix-maha}

To elaborate on the relation between RPO and the Mahalanobis distance, the latter describing an ellipsoid in the data points representation space, let us remind the results presented in~\cite{velasco2012robust}. Let us consider a data point $x \in \mathbb{R}^d$ belonging to a data matrix $X^{d \times n}$ following an Elliptically Symmetric Distribution (ESD) containing $n$ samples for which we want to compute an outlyingness score $\mathcal{O}(x)$. The ESD hypothesis guarantees that the sample covariance matrix $\Sigma$ is positive definite. Using $\Sigma$, one can define an outlyingness score using the Mahalanobis distance:

\begin{equation}
\mathcal{O}_{MAHALANOBIS}(x) = \sqrt{ (x - \mu_X)^T \Sigma^{-1} (x - \mu_X) }
\end{equation}

where $\mu_X$ is the data points mean, i.e. the mean column of $X$. According to the extended Cauchy-Schwarz inequality~\cite{johnson2014applied}, for any nonzero vector $u \in \mathbb{R}^d$:

\begin{equation}
(u^T(x - \mu_X))^2 \leq (u^T \Sigma u) ((x - \mu_X)^T \Sigma^{-1} (x - \mu_X))
\end{equation}

\begin{equation}
\label{bound-one}
\implies \dfrac{(u^T (x - \mu_X))^2}{u^T \Sigma u} \leq (x - \mu_X)^T \Sigma^{-1} (x - \mu_X)
\end{equation}

where $u^T \Sigma u \geq 0$ since $\Sigma$ is positive definite. As suggested in~\cite{johnson2014applied} if one takes $u = \alpha \Sigma^{-1} (x - \mu_X)$, with $\alpha \in \mathbb{R}^*$, one gets:

\begin{equation}
\implies \dfrac{((\alpha \Sigma^{-1}(x - \mu_X))^T (x - \mu_X))^2}{(\alpha \Sigma^{-1}(x - \mu_X))^T \Sigma (\alpha \Sigma^{-1}(x - \mu_X))} \leq (x - \mu_X)^T \Sigma^{-1} (x - \mu_X)
\end{equation}

\begin{equation}
\implies \dfrac{((\alpha \Sigma^{-1}(x - \mu_X))^T (x - \mu_X))^2}{\alpha^2 ( \Sigma^{-1}(x - \mu_X))^T \mathbb{I}_d (x - \mu_X)} \leq (x - \mu_X)^T \Sigma^{-1} (x - \mu_X)
\end{equation}

the latter leading to the equality case, showing us the upper bound is attainable. Since the upper bound is reachable for $u = \alpha \Sigma^{-1} (x - \mu_X)$, and that such format allows $u$ to be a RP of unit norm as required by the definition of RPO for Eq.~\ref{RPO}, one can intuitively conclude that using a maximum estimator over numerous RPs the bound is approached or reached in Eq.\ref{bound-one} by the left term, i.e. for $\mathbb{U}$ a large set of unit norm RPs:

\begin{equation}
\label{max-lemma}
\underset{u \in \mathbb{U}}{max} \dfrac{(u^T (x - \mu_X))^2}{u^T \Sigma u} = (x - \mu_X)^T \Sigma^{-1} (x - \mu_X)
\end{equation}

This result is actually well-known and called the Maximization Lemma in~\cite{johnson2014applied}, where instead of $u \in \mathbb{U}$, $u$ is an arbitrary nonzero vector, i.e. $\underset{u \in \mathbb{U}}{max}$ is replaced by $\underset{x \neq 0}{max}$. The proof for this lemma is actually the previous example case where $u = \alpha \Sigma^{-1} (x - \mu_X)$ is shown to make the equality case happen. Note that using the equality case format for $u$ implies knowing the covariance matrix $\Sigma^{-1}$. This emphasizes the relevance of RPO to compute a multivariate outlyingness without requiring a covariance matrix computation. From Eq.~\ref{max-lemma}, one can deduce:

\begin{equation}
\label{max-lemma-transformed}
\underset{u \in \mathbb{U}}{max} \dfrac{(u^T x - u^T \mu_X)^2}{\sigma^2(u^T x)} = (x - \mu_X)^T \Sigma^{-1} (x - \mu_X)
\end{equation}

this expression makes the projected sample $u^T x$ and the projected mean $u^T \mu_X$ appear. Note that for the denominator the transformation is as follows:

\begin{equation}
\sigma^2 (u^T x) = E \left[ (u^T(x - \mu_X)) (u^T(x-\mu_X))^T \right] = E \left[ u^T (x-\mu_X) (x-\mu_X)^T u \right] = u^T \Sigma u 
\end{equation}

Eq.~\ref{max-lemma-transformed} now only differs from RPO due to the square and the estimators of first and second order statistics: RPO replaces the mean and the variance of the projected inputs $x \in X$ with the robust estimators median and median absolute deviation respectively, which in turn adds a constant factor with respect to the Mahalanobis distance. Regarding the relevance of these robust estimators choice and the additional constant factor see~\cite{velasco2012robust}.

We thus get back to the conclusion of~\cite{velasco2012robust} indicating the equivalence (up to a constant factor) between the Mahalanobis distance and RPO computed with an infinite number of RPs under the multivariate elliptical distribution hypothesis. Whereas RPO as presented in Table~\ref{unsupervised-results} used the $max$ estimator to integrate over RPs as defined in~\cite{velasco2012robust}, Deep RPO replaces the $max$ with a $mean$ estimator in Eq.~\ref{RPO} to measure outlyingness in the latent representation space provided by a neural network. This replacement is motivated by empirical results and an interpretation provided in~\cite{bauw2021deep}. The drawback of this change is that the Mahalanobis equivalence guarantee is lost, since the Maximization Lemma leading to Eq.~\ref{max-lemma} can not be used with a $mean$. Working with an ellipsoid instead of a latent hypersphere as in~\cite{ruff2018deep} supposedly made the latent normality boundary used by the training objective more flexible and tailored to the data, and was the original motivation of~\cite{bauw2021deep}. Recall that this normality boundary is fitted to training data before training and frozen, the boundary being defined by a single location estimator in the case of Deep SVDD, and by as many location and spread estimators as there are RPs in the case of Deep RPO.

Intuitively, the $mean$ will pull the quantity integrated over the large set of RPs away from the upper bound. Even though the score is based on projected 1D outlyingnesses each normalized by their respective location and spread estimators, there is no assurance that once integrated with a $mean$ into one final outlyingness these quantities generate a normality ellipsoid similar to the Mahalanobis one. Actually, nothing indicates the integrated outlyingness describes any kind of ellipsoid in the input vector representation space. However, since the $mean$ integrates over the dimensions created by the set of RPs, and since along these dimensions each 1D coordinate is centered and normalized using its own location and spread estimators, the $mean$ still relates to an ellipsoid in the high-dimensional representation space generated by the RPs.

In order to ensure the relevance of such replacement for our application, which as previously explained in this appendix removes the equivalence with a Mahalanobis distance when numerous RPs are used to estimate RPO, Deep RPO with $mean$ and Deep RPO with $max$ were compared for every deep experimental setup presented in the paper. The resulting performances confirmed the superiority of Deep RPO with $mean$. A few of these performances are presented in Table~\ref{meanvmax-sad-ssl-results} and Table~\ref{meanvmax-experiments-contamination}. Since these experiments rely on a deep neural network, one way to see the difference between Deep RPO with $mean$ and Deep RPO with $max$ is to consider that while with $max$ the gradient will be based on a single projection, the $mean$ produces a gradient stemming from all projections simultaneously.

\begin{table}[h]
\caption{Unsupervised and semi-supervised experiments where SAD and SSL samples provide additional supervision, with Deep RPO with $max$ estimator to integrate over the RPs and Deep RPO with $mean$ (average test AUCs in \% $\pm$ StdDevs over ten seeds): the integration of 1D projected anomaly measures with a $mean$ systematically leads to better performances. Note that in these Deep RPO experiments, RPO is applied to encoded inputs in the latent representation space of the neural network.}
\label{meanvmax-sad-ssl-results}
\begin{tabular}{|l|l|l|c|c|c|c|}
\hline
AD method (input format) & SAD loss & SSL loss & Mean test AUC (1 mode) & Mean test AUC (2 modes) \\
\hline
Deep RPO (SP) - mean             & no SAD & no SSL & \textbf{83.60} $\pm$ \textbf{5.35} & \textbf{78.13} $\pm$ \textbf{6.02}\\
Deep RPO (SP) - max              & no SAD & no SSL & 74.85 $\pm$ 7.29 & 71.60 $\pm$ 6.57 \\
\hline
Deep RPO (SP) - mean  & away & away                 & \textbf{90.91} $\pm$ \textbf{5.94} & \textbf{92.69} $\pm$ \textbf{7.98} \\
Deep RPO (SP) - max   & away & away                 & 78.85 $\pm$ 9.40 & 75.53 $\pm$ 9.10 \\
\hline

\end{tabular}
\end{table}

\begin{table}[h]
\caption{Contamination experiments with Deep RPO with $max$ estimator to integrate over the RPs and Deep RPO with $mean$ (average test AUCs in \% $\pm$ StdDevs over ten seeds): except in one case the integration of 1D projected anomaly measures with a $mean$ leads to better performances. Note that in these Deep RPO experiments, RPO is applied to encoded inputs in the latent representation space of the neural network.}
\label{meanvmax-experiments-contamination}
\begin{tabular}{|l|c|c|c|c|c|c|}
\hline
AD method (input format) & SAD loss & SSL loss & Mean test AUC (1 mode) & Mean test AUC (2 modes) \\
\hline
Deep RPO (SP) - mean              & no SAD & no SSL       & \textbf{81.29} $\pm$ \textbf{5.92} & \textbf{74.82} $\pm$ \textbf{5.89} \\
Deep RPO (SP) - max               & no SAD & no SSL       & 74.21 $\pm$ 7.48 & 68.95 $\pm$ 5.09 \\
\hline
Deep RPO (SP) - mean  & no SAD & SSL c. & \textbf{86.66} $\pm$ \textbf{6.41} & \textbf{82.78} $\pm$ \textbf{8.25} \\
Deep RPO (SP) - max   & no SAD & SSL c. & 82.52 $\pm$ 7.43 & 77.29 $\pm$ 8.45 \\
\hline
Deep RPO (SP) - mean & no SAD & away         & 76.16 $\pm$ 9.87 & \textbf{76.56} $\pm$ \textbf{8.69} \\
Deep RPO (SP) - max  & no SAD & away         & \textbf{76.20} $\pm$ \textbf{9.37} & 73.41 $\pm$ 9.61 \\
\hline

\end{tabular}
\end{table}

\section{Affine invariance of RPO with max and mean estimators}
\label{appendix-invariance}

We want to prove the affine invariance of the following quantity, called the random projection outlyingness (RPO):

\begin{equation}
\label{A-RPO-max}
    O_{RPO}(x, X) = \underset{\norm{u}=1}{sup} \dfrac{\lvert u^T x - Med(u^T X) \rvert}{MAD(u^T X)}
\end{equation}

that is, we want to prove the following equality:

\begin{equation}
\label{A-RPO_affinv}
    O_{RPO}(Ax+b, AX+b) \stackrel{?}{=} O_{RPO}(x, X)
\end{equation}

where:

\begin{itemize}
    \item $x \in \realset^{d\times1}$ is the data point for which we want to compute an outlyingness measure;
    \item $X \in \realset^{d \times n}$ is the data matrix containing $n$ features vectors in $\realset^d$ (i.e. the data distribution, including $x$);
    \item $u \in \realset^{d\times1}$ is a random projection vector of unit norm, \emph{i.e.} $u \in S^{d-1}$ where $S^{d-1} = \{x \in \mathbb{R}^d:\norm{x}_2=1\}$;
    \item $A \in \realset^{d \times d}$ is a non-singular matrix (for the affine transformation);
    \item $b \in \realset^{d \times 1}$ is a constant vector (for the affine transformation);
    \item $Med(u^T X)$ is the median of the scalars generated by the 1D projection of all $x$ in $X$ by $u$;
    \item $MAD(u^TX)$ is the median absolute deviation of the same scalars, \emph{i.e.}\\ $MAD(u^TX) = Med(\lvert u^Tx - Med(u^TX) \rvert)$.
\end{itemize}

$AX+b$ is a permissive notation defining the affine transformation of every column features vector with $A$ and $b$, \emph{i.e.} the affine transformation of the whole data distribution on which the location and scatter estimators, respectively the $Med$ and the $MAD$, are applied. We also want to prove the affine invariance of that same quantity where the $sup$ is replaced by a $mean$ estimator:

\begin{equation}
\label{A-RPO-mean}
    O_{RPO-MEAN}(x, X) = \underset{\norm{u}=1}{mean} \dfrac{\lvert u^T x - Med(u^T X) \rvert}{MAD(u^T X)}
\end{equation}

This RPO-MEAN is the RPO variant introduced in~\cite{bauw2021deep} and used in the encoding neural network output space for our Deep RPO experiments. The Eq.~\ref{A-RPO-max} stems from~\cite{donoho1992breakdown,huber1985projection} and defines the outlyingness used to generate the statistical depth~\cite{zuo2000general} called random projection depth (RPD), the latter corresponding to the following expression: $RPD(x, X) = \frac{1}{1 + O_{RPO}(x, X)}$. Note that this depth, and the associated outlyingness of Eq.~\ref{A-RPO-max}, are defined with an infinite number of random projections $u \in \realset^d$, thus the two quantities (RPO and RPD) can only be implemented with a stochastic approximation, \emph{i.e.} with a large but finite number of random projections. For instance, in the experiments described in this paper, the $sup$ and $mean$ over all RPs of unit norm are approximated with a $max$ and $mean$ over 1000 RPs respectively. For other experiments on RPO with diverse quantities of RPs, see~\cite{bauw2021deep}.


\paragraph{Proof:}

Let us start by noticing that both the upper and lower parts of the RPO ratio are invariant to the bias term $b$ of the affine transformation:

\begin{align}
    \lvert u^T (Ax +b) - Med(u^T (AX+b)) \rvert &= \lvert u^T Ax + u^Tb - Med(u^T AX) - u^Tb \rvert \\
    &= \lvert u^T Ax - Med(u^T AX) \rvert \\
    MAD(u^T (AX+b)) &= Med( \lvert u^T (Ax+b) - Med(u^T (AX+b)) \rvert) \\
    &= Med( \lvert u^T Ax + u^Tb - Med(u^T AX) - u^Tb \rvert) \\
    &= Med( \lvert u^T Ax - Med(u^T AX) \rvert) \\
    &= MAD(u^T (AX))
\end{align}

This indicates both RPO and RPO-MEAN are translation invariant, a partial requirement to achieve affine invariance. Let us factor the upper and lower parts of the RPO ratio to make a unit norm vector $\dfrac{u^T A}{\norm{u^T A}}$ appear:

\begin{align}
    \lvert u^T (Ax) - Med(u^T (AX)) \rvert &= \left| \norm{u^T A} \dfrac{u^TA}{\norm{u^T A}} x - Med \left( \norm{u^T A} \dfrac{u^T A}{\norm{u^T A}} X \right) \right| \\
    &= \norm{u^T A} \left| \dfrac{u^TA}{\norm{u^T A}} x - Med \left( \dfrac{u^T A}{\norm{u^T A}} X \right) \right| \\
    MAD(u^T (Ax)) &= Med \left( \left| \norm{u^T A} \dfrac{u^TA}{\norm{u^T A}} x - Med \left( \norm{u^T A} \dfrac{u^TA}{\norm{u^T A}} X \right) \right| \right) \\
    &= \norm{u^T A} Med \left( \left| \dfrac{u^TA}{\norm{u^T A}} x - Med \left( \dfrac{u^TA}{\norm{u^T A}} X \right) \right| \right) \\
\end{align}

This enables us to rewrite the RPO ratio as follows:

\begin{align}
    \dfrac{\lvert u^T Ax - Med(u^T AX) \rvert}{MAD(u^T AX)} &= \dfrac{\norm{u^T A} \left| \dfrac{u^TA}{\norm{u^T A}} x - Med \left( \dfrac{u^T A}{\norm{u^T A}} X \right) \right|}{\norm{u^T A} Med \left( \left| \dfrac{u^TA}{\norm{u^T A}} x - Med \left( \dfrac{u^TA}{\norm{u^T A}} X \right) \right| \right)} \\
    &= \dfrac{\left| \dfrac{u^TA}{\norm{u^T A}} x - Med \left( \dfrac{u^T A}{\norm{u^T A}} X \right) \right|}{Med \left( \left| \dfrac{u^TA}{\norm{u^T A}} x - Med \left( \dfrac{u^TA}{\norm{u^T A}} X \right) \right| \right)}
\end{align}

Thus for any $u$, $x$ and $X$, let $f(u)$ be:

\begin{equation}
	f(u) := \dfrac{\lvert u^T x - Med(u^T X) \rvert}{MAD(u^T X)}
\end{equation}

if one defines $\phi(u) := \dfrac{u^TA}{\norm{u^TA}}$ and $\psi(u) := u^T A$, we have $f \circ \phi(u) = f \circ \psi(u)$. Moreover, since $\phi$ is a bijection from $S^{d-1}$ to $S^{d-1}$, for $g$ the $mean$ or $sup$ operator applied to every existing random projection $u$:

\begin{equation}
    \underset{u \in S^{d-1}}{g} \left[ f(u) \right] = \underset{u \in S^{d-1}}{g} \left[ f \circ \phi(u) \right]
\end{equation}

Combining the two last equalities provides us with the invariance to the linear transformation defined by $A$:

\begin{equation}
    \underset{u \in S^{d-1}}{g} \left[ f(u) \right] = \underset{u \in S^{d-1}}{g} \left[ f  \circ \psi(u) \right]
\end{equation}

In other words, since $\phi$ is a bijection and all existing random projections of unit norm are considered during the integration over $S^{d-1}$, the operator $g$ is not affected whether the RPs are transformed by $\phi$ beforehand or not. The intuition behind this invariance is that we are looking over the same infinite set of random projections, the transformation matrix $A$ at most reshuffling the RPs in the infinite set the estimator integrates over. More generally, that is true for any permutation invariant\footnote{An operator is permutation invariant if any permutation of the inputs does not change the output: \begin{equation*}
g(x_1,~...~, x_t) = g(x_{\pi(1)},~...~, x_{\pi(t)})
\end{equation*} for any permutation $\pi$.} operator $g$.

The previous translation invariance with respect to $b$ and the linear transformation invariance with respect to $A$ prove the affine invariance of both RPO and RPO-MEAN.

\end{document}